# Long-term Effects of India's Childhood Immunization Program on Earnings and Consumption Expenditure: Comment

David Roodman

June 5, 2024

**Abstract**
Summan, Nandi, and Bloom (2023), hereafter SNB, find that exposure during infancy to India's Universal Immunization Programme (UIP) increased wages and per-capita household expenditure in early adulthood. SNB regress these outcomes on a treatment indicator that depends upon year and district of birth while controlling for age at follow-up. Because year of birth and age are nearly collinear, SNB's identifying variation does not come from the staggered introduction of the UIP, but rather from the progression of time during the follow-up period. Within the 12-month follow-up period, those interviewed later were more likely to have been treated and, on average, reported higher wages and household expenditure. Wages and household expenditure, however, rose by at least as much in a control group composed of people too old to have been exposed as infants to the UIP as in the treated group. SNB's results are best explained by inflation, economic growth, and non-random survey sequencing during the follow-up survey period.

# I. Introduction

Summan, Nandi, and Bloom (2023) study the long-term effects of India's Universal Immunization Programme (UIP). The authors (hereafter, SNB) find that infants exposed to the UIP in the late 1980s had higher wages in early adulthood than their unexposed counterparts (0.138 log points) and higher per-capita household expenditure (0.028 log points). SNB conclude that there is "a direct link between vaccination and labor market outcomes" (SNB 2023, p. 574).

In this comment, I provide evidence that SNB's conclusion is unwarranted. SNB construct their treatment variable from the staggered rollout of the UIP in the late 1980s across districts. SNB classify individuals as having been exposed to the UIP if they are estimated to have been born in a year and district in which the UIP had begun operating. While regressing their outcomes on UIP exposure, , which is based on year of birth, SNB control for age at follow-up. Because year of birth and age at follow-up are nearly collinear, SNB's identifying variation is not coming from the staggered introduction of the UIP.

To demonstrate the source of SNB's identifying variation, I conduct a randomization exercise in which the program's launch is hypothetically and arbitrarily rescheduled across districts within the rollout period, 1986–90. Running SNB's regression specifications on 100,000 treatment-randomized data sets produces large, positive, and statistically significant estimates. Randomized UIP treatment is on average associated with a 0.088 log point increase in the wages and a 0.038 log point increase in per-capita household expenditure. These "design effects" can explain most of the wage impact found by SNB and all of the household expenditure impact.

If the precise timing of treatment in each district is nearly immaterial to the SNB results, then where do the results come from? Because of how the two nearly-collinear variables, birth year and age, are constructed, identification ultimately derives from the progression of time during the 12 months of the follow-up survey in 2011–12. Indeed, when one examines the data in this frame, one finds the essence of SNB's results. Later-interviewed people of a given age on the day of interview are more likely to be deemed treated and to report higher wages and household expenditure.

Since variation in survey time is the main locus of identification, the SNB findings are susceptible to confounding from trends in survey time that have little to do with vaccination. The wages and expenditures reported by interviewees may have risen during the survey because of inflation, or economic growth, or non-random sequencing of interviews. If such factors are the true source of SNB's results, then there should have been comparable increases in the wages and household expenditures among interviewees born too early to have benefited from the UIP. This is precisely what I find. It is therefore likely that forces other than vaccination explain SNB's results.

**II. Background: SNB's data, methods, and results**

India launched the UIP In 1985. The goal was ambitious: to immunize all pregnant women and 85% of infants against measles, polio, and four other diseases (Lahariya 2014). The program started in 31 of India's districts in fiscal year 1985–86 and expanded to all 466 by 1989–90. Previous studies find that the UIP increased height and weight among children under 4 and boosted schooling attainment over the course of childhood (Anekwe and Kumar 2012; Nandi et al. 2020).

SNB study the long-term economic effects of UIP exposure without actually observing whether an infant was immunized. Instead, they consider an individual, $i$, to have been immunized if the UIP had reached the individual's district before their birth. As an example, if the UIP is recorded as having started in a district during fiscal year 1985–86, which began on April 1, SNB consider individual $i$ to have been treated if they were born in that district in or after 1986. SNB estimate date of birth as date of interview minus the number of completed years of life as of the interview.

SNB's outcomes come from the 68$^{th}$ National Sampling Survey (NSS), which was in the field from July 1, 2011 through June 30, 2012. The NSS data files are publicly available from India's National Data Archive and provide information on wages, household size, and household expenditure.[1] The NSS also contains demographic and educational attainment variables, some of which SNB use as controls. In their base specifications, SNB restrict samples to NSS respondents who were 21 through 26 years old when interviewed; the subjects were therefore born between 1985 and 1991, a range that just encompasses the UIP rollout.

---

[1] microdata.gov.in/nada43/index.php/catalog/126/get_microdata.

Given the staggered introduction of the treatment and the single, cross-sectional follow-up, a typical research design would be two-way fixed effects (TWFE). Indeed, the paper's prominent map showing when each district first received the UIP could lead the reader to assume that SNB estimate a standard TWFE model. The estimating equation would be:

$$(1)\ Y_{ijt} = \delta D_{jt} + \boldsymbol{\beta}'\mathbf{C}_{ijt} + \nu_j + \eta_t + \epsilon_{ijt},$$

where $Y_{ijt}$ is an outcome for individual $i$ born in district $j$ in year $t$, $D_{jt}$ is a dummy for UIP exposure, $\mathbf{C}_{ijt}$ is a vector of individual-level controls, the $\nu_j$ are birth district fixed effects, and the $\eta_t$ are birth year fixed effects.

SNB do not, however, estimate a TWFE regression model. Instead, SNB estimate an equation closer to:

$$(2)\ Y_{ijt} = \delta D_{jt} + \boldsymbol{\beta}'\mathbf{C}_{ijt} + \lambda_{jt} + \epsilon_{ijt},$$

in which the district-of-birth and year-of-birth fixed effects are replaced with district-by-year fixed effects, $\lambda_{jt}$. In fact, equation (2), is not identified, because the treatment $D_{jt}$ is collinear with the fixed effects, $\lambda_{jt}$. SNB actually estimate:

$$(3)\ Y_{ijt} = \delta D_{jt} + \boldsymbol{\beta}'\mathbf{C}_{ijt} + \lambda_{jt'} + \epsilon_{ijt},$$

where $\lambda_{jt}$ has been replaced with $\lambda_{jt'}$, and $t'$ is the respondent's age in completed years on the day of interview. In effect, SNB leverage year of birth ($t$) to identify the impact of UIP exposure while controlling for age in 2011–12 ($t'$). This design may be counterintuitive, but SNB are clear about using it. They write: "The source of variation at the individual level is from the year of birth, controlling for district and age" (SNB 2023, p. 563). Equation (3) is technically identified because $t$ and $t'$ are not quite collinear. As detailed below, in section IV, some subjects have the same $t$ but different $t'$, or vice versa.

The first column of Table 1 reproduces SNB's preferred estimates of $\delta$ in equation (3). The results in column (2) are generated with data and code sent me by the SNB corresponding author, Amit Summan. These nearly match the published results and are my best replication thereof. UIP exposure as an infant is associated with a 0.129 log point increase in wages among 21- though 26-year-olds, and a 0.028 log point increase in household expenditures.[2]

In reconstructing the SNB data set from primary sources, I discovered some errors in the original data set. They are small and do not greatly affect results. In the appendix, I present estimates based on the corrected data.

**III. Applying the NSS survey weights**

The 68th NSS employed a complex survey design. In the second sampling stage, households were split into three strata based on per-capita household expenditure. The strata were over- or under-sampled relative to the subpopulations they represented.[3] The NSS data set includes a weight variable that researchers can use to adjust for the disproportionate sampling. When a subject's probability of inclusion in a regression sample depends on or is related to the outcome of interest, unweighted estimates are in general biased (Hausman and Wise 1981; Solon, Haider, and Wooldridge 2015). However, weighting increases the variance of linear regression. This is especially the case when, as in the NSS, the distribution of weights is highly skewed. In the sample for the SNB log wage regression in column (2) of Table 1, the median weight is 2800 but the 99th percentile is 22,000. To reduce the influence of extreme weights, I trim high values to the median weight plus 5 times the interquartile range (Potter and Zheng 2015).

In column (3) of Table 1, I report estimates of equation (3) incorporating the NSS survey weights. The wage impact estimate drops modestly, from 0.129 to 0.119 log points, while the consumption impact rises modestly, from 0.028 to 0.040. Although weighting increases the standard errors for both by about 50 percent, the estimates remain statistically significant at conventional levels.

**IV. The source of identification in SNB**

Above, I observe that 1) SNB define treatment status based on an individual $i$'s year and place of birth; and 2) SNB control for $i$'s age when surveyed, which is nearly collinear with year of birth. It is not immediately obvious what these two statements, taken together, imply for the interpretation of the SNB results. This section delves into that question, first examining where SNB's identification does *not* come from, then where it *is* coming from.

---

[2] All data and code for this comment are at doi.org/10.7910/DVN/ECICGT.
[3] See icssrdataservice.in/datarepository/index.php/catalog/92/export.

Although the staggered rollout of the UIP is the basis of the treatment definition in SNB, it is not the source of identification. To demonstrate this claim, I perform a sort of randomization inference. In general, randomization inference entails arbitrarily and repeatedly reassigning treatment in an already-gathered data set, in order to determine the distribution of an estimator on that data set (Fisher 1935; Heß 2017; Young 2018). Here, I scramble the map of which districts first received the UIP when, while preserving the number of districts that received it in each fiscal year. I do so 100,000 times. After each scrambling, I recompute each individual's treatment status and rerun the unweighted and weighted regressions in columns (2) and (3) of Table 1. I plot the distributions. And I compute their means, which I call "design effects." See Figure 1.

The design effects are closer to SNB's results than to zero, and so can largely account for the SNB results. In the unweighted SNB wage specification (upper left of Figure 1), the average "impact" of the randomized treatment on log wages is 0.088 points. That is marked in red while the SNB point estimate of 0.129 is marked in blue. Under the randomization distribution, the two-tailed $p$ value for the null hypothesis that the SNB point estimate equals the design effect is 0.14. For household expenditures, the SNB estimate of 0.028 falls *below* the randomization mean of 0.038 ($p$=0.23; upper right of Figure 1). Weighting observations in the regressions increases the variance of the simulations and modestly raises the design effects relative to original-sample results (lower half of the figure). As documented in Figure A1 in the appendix, introducing the data corrections brings the design effects and full-sample point estimates closer together and raises all the $p$ values.

How can a randomized treatment yield regression coefficients that are, on average, well above zero? The answer lies in the details of the randomization. In all iterations, 0% of the oldest subjects—those born in 1985—are classed as treated while 100% of the youngest—born in 1991—are. More generally, no matter how the rollout map is scrambled, the probability that a subject is treated rises with birth year. If the randomization were done at the individual level, with each person's treatment status set to 0 or 1 in a way unrelated to birth year, this would not have been the case. The upshot for SNB is that their results largely derive from treatment variation associated with year of birth, since that is what is preserved in the randomization.

However, much of the variation in year of birth is absorbed by the controls—notably, age at the time of interview. For insight into how controlling for age affects identification, imagine two natives of the district of Anantapur in Andhra Pradesh.

According to the SNB data, Anantapur first received the UIP in fiscal year 1985–86. Anantapurians born in 1986 or later are therefore taken as young enough to have been treated. Suppose that one of the Anantapur natives is interviewed on the first day the follow-up survey is fielded, July 1, 2011, and reports being 26 years old on that day. The SNB methodology would estimate that the person was born in 2011–26 = 1985, and classify the person as untreated. Suppose that the other person is contacted at the end of the survey a year later, and also reports being 26. This person would be estimated to have been born in 1986 rather than 1985—and so be classified as treated, because the UIP had just begun operating then. These two respondents have different birth years, but the same age at survey and same birth district. They therefore illustrate how there can be variation in birth year while controlling for age and birth district. Suppose finally that the later-interviewed subject is not only more likely to be treated, abut also earns and spends more. Then we would observe that the treatment and the outcomes are positively related, even as we hold fixed (i.e., control for) the age at interview.

The SNB analysis should be seen as a scaled-up version of this comparison. That is to say, the identifying variation in SNB is associated with the progression of time during the follow-up survey. Later-interviewed people of a given age are more likely treated and evidently have better outcomes on average.

**VI. Effects in survey time**

One way to test the claim that the progression of time during the survey is the locus of SNB's identification is to check how removing most of that variation affects their results. To that end, I add survey month dummies to the control set $\mathbf{C}_{ijt}$ in equation (3) and report the regression results in columns (4) and (5) of Table 1. The estimated impact on wages falls by a factor of 2–3 and that on consumption disappears. The average *t* statistic for these four new results is –0.14, which is consistent with the estimates being randomly dispersed around 0. Because little treatment variation remains, I do not view the new results as meaningful impact estimates. They serve, rather, to confirm that the progression of survey time is central to identification in SNB.

Now, accepting that centrality does not in itself invalidate SNB's conclusions. If, because of nationally rising vaccination rates in the late 1980s, people of a given reported age who were interviewed later in the 2011–12 follow-up were more treated on average, as in the Anantapur example; and if because of vaccination they earned and spent more; that could generate the SNB results and validate SNB's interpretation.

However, a sharper understanding of the source of the results also opens the door to a sharper analysis of identification issues, such as confounding. If non-UIP factors lifted wages and expenditure during the survey period, the SNB design could wrongly attribute those effects to the UIP. Two potential confounders are inflation and economic growth. Between mid-2011 and mid-2012, consumer prices in India rose 10.6%. Gross domestic product per capita increased 13.3% in nominal rupee terms.[4] In addition, survey effects could be at work: while the NSS used a stratified random sample, the *sequence* of the sampling—which villages and neighborhoods were surveyed when—was not necessarily random.

To investigate these ideas, Figure 2 plots key variables over the survey period. Moving averages, using an Epanechnikov smoothing kernel with a bandwidth of 30 days, are drawn in black. Best-fit lines are in green. As in SNB, survey weights are not incorporated. The leftmost plot confirms that in the main SNB sample—people born in 1985–90—the treated fraction rises during the survey period, approximately from 0.39 to 0.43. Most of the increase occurs around January 1, 2012, because of how SNB estimates birth year from age and survey date. The middle and right plots show that the outcomes of interest also rise in survey time. Among people born in 1985–90, wages climb by 0.2 log points and per-capita household expenditure by 0.1 log points.

One component of those climbs, inflation, can be directly removed from the SNB regressions. Columns (6) and (7) of Table 1 show the effect of adjusting the outcomes for inflation. (These regressions do not include survey month dummies.) Compared to columns (2) and (3), all point estimates drop by about 0.03 log points. When incorporating survey weights, the impact estimates are 0.086 log points for wages and 0.009 for consumption. The wage result is still reasonably distinguishable from zero, but not the consumption result. Inflation, then, does not appear to generate all of the wage impact in SNB, but can explain the consumption impact.

---

[4] Figures based on the monthly series, "Prices, Consumer Price Index, All items," the quarterly "National Accounts, Expenditure, Gross Domestic Product, Nominal, No Seasonal Adjustment, Domestic Currency," and the annual "Population, Persons, Number of," from the International Monetary Fund, International Finance Statistics, data.imf.org, accessed May 13, 2024. GDP growth is based on the geometric average of the Q2 and Q3 year-over-year expansions.

We cannot be certain about how much the inflation-adjusted results owe to vaccination, and how much to confounding from economic growth and non-random survey sequencing. But we can approach the question indirectly. If confounders dominate, we should expect wages and consumption to rise about as much during the survey in other age groups—in particular, in a group of respondents too old to have been exposed to the UIP in infancy. We should expect the opposite if the UIP is the main causal factor.

To check these competing predictions, Figure 2 includes trends for interviewees who were born in 1975–80, 10 years before those in the main SNB sample. The pattern is stark: while the control group's ascribed treatment rate held at zero throughout the survey period (left pane), wages rose almost exactly as much in the control group as in the treatment group, at 0.219 versus 0.217 log points under the linear fits; and per-capita household spending rose faster in the control group, at 0.122 log points versus 0.094. These findings strongly favor non-UIP factors as the source of the SNB results.

**VII. Conclusion**

Although SNB base their treatment variable on the staggered rollout of India's vaccination program in the late 1980s, the identifying variation in their results arises elsewhere. In a randomization exercise, the SNB impact estimator averages 0.088 for log wages and 0.038 for log per capita household expenditure. These design effects can explain most of SNB's actual estimate of 0.129 for wages and more than explain SNB's consumption estimate of 0.028. Correcting modest data errors brings the design effects and original-sample estimates even closer together.

Understanding the source of the SNB results therefore requires understanding the source of these design effects. What the randomization exercise preserves is that, on a national basis, the treated fraction rises with birth year. SNB's identifying variation is therefore largely restricted to variation associated with birth year.

The near-collinearity between birth year and a control, age at interview, further narrows the identifying variation, to the progression of time during the 12 months of the follow-up survey. Within SNB's sample, people interviewed later were more likely to be deemed treated, and they reported higher wages and household expenditures, in nominal rupees. Secular factors such as inflation, economic growth, and non-random survey sequencing could explain the improving outcomes during the survey.

And they likely do, since outcomes improve at least as much for people too old to have received UIP treatment. SNB's conclusion that vaccination in infancy improved economic outcomes in early adulthood is not supported by the evidence they present.

**Table 1. Estimated impact of UIP on wages and expenditures, those born 1985–90**

| Outcome | Published | SNB data & code | | Revised, survey month effects | | Revised, inflation-adjusted | |
|---|---|---|---|---|---|---|---|
| | (1) | (2) | (3) | (4) | (5) | (6) | (7) |
| Observation weights? | No | No | Yes | No | Yes | No | Yes |
| Log wage | 0.14 | 0.129 | 0.119 | 0.061 | 0.037 | 0.097 | 0.086 |
| | (0.03) | (0.029) | (0.047) | (0.034) | (0.054) | (0.029) | (0.047) |
| Observations | 10,781 | 10,781 | 10,781 | 10,781 | 10,781 | 10,781 | 10,781 |
| Log monthly per- | 0.03 | 0.028 | 0.040 | −0.015 | −0.025 | −0.003 | 0.009 |
| capita expenditure | (0.01) | (0.011) | (0.016) | (0.010) | (0.016) | (0.011) | (0.016) |
| Observations | 46,557 | 46,557 | 46,557 | 46,557 | 46,557 | 46,557 | 46,557 |

Results in column 1 copied from upper left corners of SNB Tables 2 and 3. Results in column 2 are generated from data and code provided by the SNB corresponding author. Column 3 adds survey weights. In columns 4 and 5, monthly fixed effects are added as controls. Regressions in columns 6 and 7 instead adjust for inflation using monthly consumer price index from the International Monetary Fund's *International Financial Statistics*. Standard errors in parentheses, clustered by district.

# Figure 1. Distribution of impact estimate under randomization of treatment

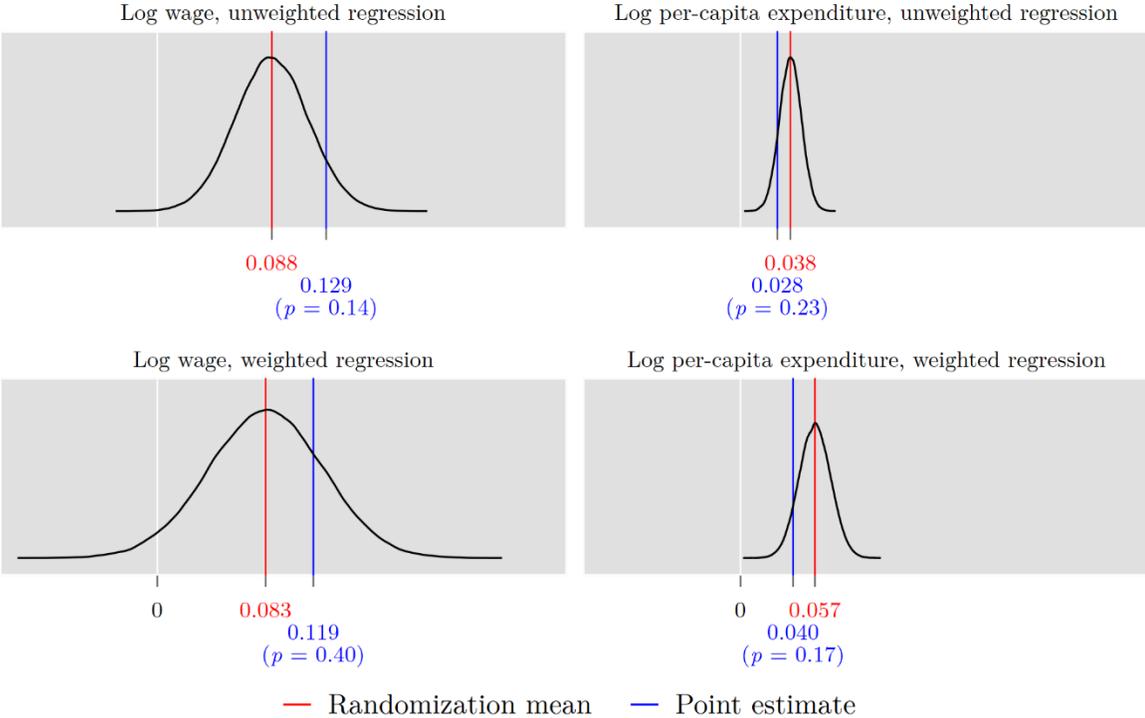

Each plot corresponds to a regression result reported in Table 1, column 2 or 3. The black curves depict the distribution of the estimate when randomly scrambling when, within 1986–90 , the vaccination program arrived in each district. The vertical red lines show the "design effects," i.e., the means of the distributions. The blue lines show the original point estimates. The displayed $p$ values are for the two-tailed test under these distributions that the point estimates differ from the design effects. 100,000 simulations are run for each plot.

**Figure 2. Treatment status, log wages, and log per-capita expenditures versus survey date: linear and locally smoothed fits**

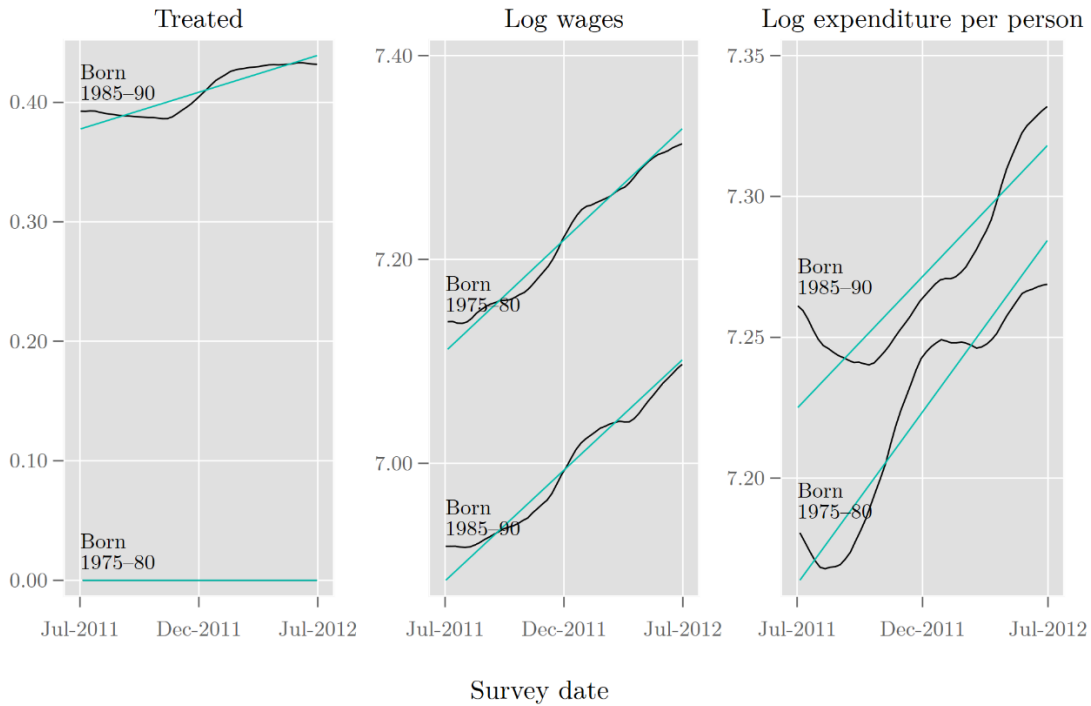

Black curves are moving averages of the indicated variables with respect to survey date, from India's National Sampling Survey of 2011–12. An Epanechnikov kernel with a bandwidth of 30 days is used. Green lines are linear fits. Fits are computed separately for the sample for key SNB regressions—those born between 1985 and 1990—and for those born between 1975 and 1980.

## Acknowledgements
Thanks to Amit Summan for detailed feedback, to Otis Reid and Michael Wiebe for comments.

**Appendix: Data corrections**

Reconstructing the SNB data set from primary sources led to the discovery of a few errors in the original. These do not greatly affect results, and so are not discussed in the main text. The errors fall under these headings:

- *Mapping districts of the late-1980s to districts of 2011–12.* Linking the SNB treatment indicator—which districts first got UIP when—to the follow-up data set is complicated by the tendency of Indian districts to split and merge over time. I independently construct a concordance between districts of the late 1980s and the districts of the 2011–12 follow-up survey. I rely primarily on Kumar and Somanathan (2016). The new concordance agrees with the SNB concordance for 595 of the 627 districts of 2011–12. The main source of discrepancy is the handling of "multi-parent districts," ones that combine territory from more than one older district. Because of ambiguity, I code the treatment variable as missing if a district has "mixed parentage," meaning that its parents did not all receive the UIP program year on the same schedule. More precisely, I code treatment as missing if the population-weighted standard deviation of the UIP treatment-arrival years of a district's parents exceeds a *de minimus* 0.1. I also correct a few apparent errors in the SNB treatment variable and its mapping to 2011–12 districts. There are two districts named Bilaspur, leading to mistaken handling in SNB of Bilaspur, Himachal Pradesh. Ashoknagar, Madhya Pradesh, is a "child" of Guna, not Gwalior. Raigarh, Maharastra, usually known as "Raigad," appears to have been confused with Raigarh district in Chhattisgarh. In addition, I mark the treatment status for Bongaigaon, Chirang, and Goalpara in Assam as missing because I believe they are children of the original Goalpara district, for which I do not see a treatment year in the SNB data.

- *Fixed-effect singletons.* The SNB samples contain some 100–200 fixed effect *singletons*—observations that are the sole instance of their combination of age and district of birth. Correia (2015) warns that retaining singletons can distort clustered standard errors. Dropping them does not affect point estimates.
- *Variable construction.* A few observations with missing marital status in the primary source are coded in SNB as unmarried, but here are treated as missing. One of the SNB controls is an estimate of the probability that a person is a migrant; it is based on an initial fitting of a probit model to separate data, the constant term from which is erroneously dropped when computing the migrant probabilities.

Table A1 and Figure A1 are constructed just like Table 1 and Figure 1 in the main text, except that the data corrections are incorporated everywhere but in the first column of the table.

**Table A2. Estimated impact of UIP on wages and expenditures, those born 1985–90, with data corrections**

| Outcome | Published | SNB data & code | | With survey month fixed effects | | Revised, inflation-adjusted | |
|---|---|---|---|---|---|---|---|
| | (1) | (2) | (3) | (4) | (5) | (6) | (7) |
| Observation weights? | No | No | Yes | No | Yes | No | Yes |
| Log wage | 0.14 | 0.115 | 0.108 | 0.046 | 0.030 | 0.083 | 0.075 |
| | (0.03) | (0.028) | (0.048) | (0.033) | (0.053) | (0.028) | (0.048) |
| Observations | 10,781 | 9,800 | 9,800 | 9,800 | 9,800 | 9,800 | 9,800 |
| Log monthly per- | 0.03 | 0.030 | 0.052 | −0.007 | −0.004 | −0.002 | 0.021 |
| capita expenditure | (0.01) | (0.010) | (0.015) | (0.010) | (0.015) | (0.010) | (0.015) |
| Observations | 46,557 | 46,352 | 46,352 | 46,352 | 46,352 | 46,352 | 46,352 |

This table is the same as Table 1 in the main text except that data corrections are incorporated in all but the first column.

**Figure A3. Distribution of impact estimate under randomization of treatment**

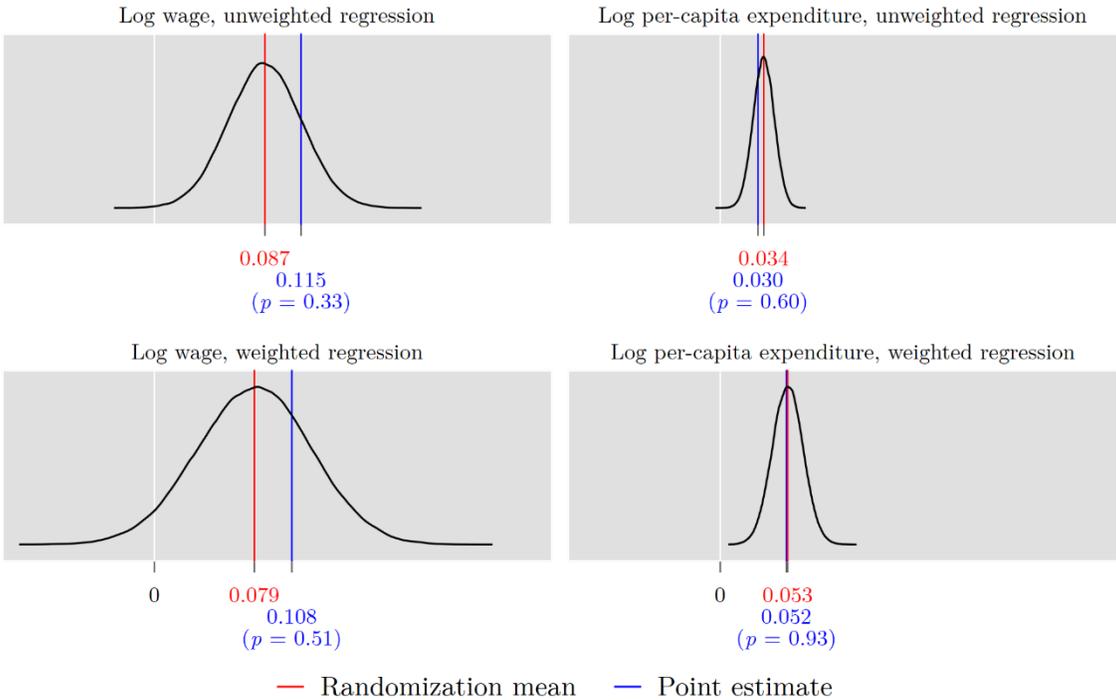

This figure is the same as Figure 1 in the main text, except that data corrections are incorporated throughout.